 \documentclass[iop]{emulateapj}

\shorttitle{M31 ULX2 with {\em Chandra}, {\em HST} and {\em XMM-Newton}}
\shortauthors{Barnard et al.}

\begin{document}


\title{The second ULX transient in M31: {\em Chandra}, {\em HST} and XMM observations, and evidence for an extended corona}


\author{R. Barnard, and M. Garcia}
\affil{Harvard-Smithsonian Center for Astrophysics, Cambridge MA 02138}
\and
\author{S. S. Murray}
\affil{Johns Hopkins University, Baltimore, Maryland, and CFA}


\begin{abstract}
XMMU\thinspace J004243.6+412519 is a transient X-ray source in M31, first discovered 2012 January 15. Different approaches to fitting the brightest follow-up observation gave luminosities 1.3--2.5$\times 10^{39}$ erg s$^{-1}$, making it the second ultraluminous X-ray source (ULX) in M31,  with a probable black hole accretor. These different models represent different scenarios for the corona: optically thick and compact, or optically thin and extended. We obtained {\em Chandra} ACIS and {\em HST} ACS observations of this object as part of our transient monitoring program, and also observed it serendipitously in a 120 ks {\em XMM-Newton} observation. We identify an optical counterpart at J2000 position 00:42:43.70 +41:25:18.54; its F435W ($\sim$ B band) magnitude was 25.97$\pm$0.03 in the 2012 March 7 observation, and $>$28.4 at the 4$\sigma$ level during the 2012 September 7 observation, indicating a low mass donor. We created two alternative lightcurves, using the different corona scenarios, finding linear decay for the compact corona and exponential decay for the extended corona; linear decay implies a disk that is $>$5 magnitudes brighter than we observed. We therefore favor the extended corona scenario, but caution that there is no statistical preference for this model in the  X-ray spectra alone.  Using two empirical relations between the X-ray to optical ratio and the orbital period, we estimate a period of $\sim$9--30 hr; this period is consistent with that of the first ULX in M31 (18$^{+5}_{-6}$ hr).
\end{abstract}





\section{Introduction}

Ultra-luminous X-ray sources (ULXs) are point sources with X-ray luminosities exceeding the Eddington limit for a 10 M$_{\odot}$ black hole ($\sim$1.3$\times 10^{39}$ erg s$^{-1}$) that are unrelated to galaxy nuclei \citep[see e.g.][for reviews]{roberts07, feng11}. ULXs are often associated with regions of high star formation such as spiral arms of galaxies, and therefore many are likely to be related to high mass X-ray binaries \citep[see e.g.][and references within]{mapelli10,mineo12}. Some ULXs may contain intermediate mass black holes \citep[ESO 243-49   HLX-1 being a strong candidate,][]{farrell09}. Other ULXs may harbor stellar mass black holes, exhibiting true super-Eddington accretion \citep[see e.g.][]{gladstone09,barnard10}, or apparent super-Eddington luminosities due to beaming \citep[e.g.][]{king01}. Since ULXs are simply defined by their observed luminosities, their membership is likely to include a wide range of diverse systems.

High quality  ULX spectra are often described by a two-component emission model, consisting of a disk blackbody component, and a component such as {\sc comptt} that represents inverse-Comptonization of cool photons on hot electrons in an accretion disk corona. Some people expect the temperature of the seed photons ($T_0$) to equal the inner disk temperature (k$T_{\rm in}$), assuming that the corona is compact and so only sees the inner disk \citep[see e.g.][]{roberts05,goncalves06}. However, many high quality ULX spectra reject models where $T_0$ = k$T_{\rm in}$ \citep{gladstone09}; instead, successful models fall into one of two types. 

The first type results in cool, optically thick coronae where $T_0$ $>$ k$T_{\rm in}$;  \citet{gladstone09}  proposed that these spectra represent a separate ``ultra-luminous state'' where an opaque corona hides the inner disk from view, meaning that the measured value for k$T_{\rm in}$ comes from further out in the disk, outside the corona. ULXs fitted this way tend to have a soft excess, and a roll-over in the spectrum above 3 keV \citep{gladstone09}. This state is distinct from the spectral states observed in Galactic X-ray binaries, where the optical depth is $\la$1. The corona must be  compact, so that we still see the inner disk in X-rays; this is also different from Galactic X-ray binaries, where corona estimates range over $\sim$20,000--700,000 km for high inclination systems \citep{church04}, while Chandra grating spectroscopy of Cygnus X-2  suggests a corona $\sim$100,000 km \citep{schulz09}.

 The second type results in optically thin coronae with $T_0$ $<$ k$T_{\rm in}$ and is also well described by a disk blackbody + power law model for some sources; these spectra represent an extended corona \citep{barnard10}, and may be related to the steep power law state observed in canonical black hole binaries \citep[see e.g.][]{remillard06}. The electron temperature determines whether or not the Comptonized component is well described by a power law; if the electron temperature is too close to (or within) the observed energy band, then the power law fails and more specialized models of Comptonization are required (e.g. {\sc comptt}). These coronae are optically thin, meaning that we are able  to see a portion of the disk emission even though when the  disk is extended. Extended coronae allow super-Eddington emission while keeping the local accretion rate sub-Eddington.
 
When disk-blackbody + power law models are applied, the power law component sometimes  dominates the low energy emission; this has regarded as unphysical by some \citep[see e.g.][]{roberts05,goncalves06}. This led  \citet{steiner09}  to   develop the {\sc simpl} convolution model  of Comptonization for XSPEC v.12 that results in a single, Comptonized emission component. \citet{barnard10} modeled {\em XMM-Newton} emission spectra of three ULXs in NGC253 (plus the dynamically confirmed  BH+Wolf-Rayet binary IC10 X-1) with two component models and convolution models. The convolution model was rejected for NGC253 ULX3 and IC10 X-1 (and the luminosity for IC10 X-1 was well sub-Eddington). NGC253 ULX1 and NGC253 ULX2 preferred the two component model over the convolution model (with lower $\chi^2$ for the same number of degrees of freedom). We infer from these results that the soft excess is real, and favor an extended corona (i.e. $T_0$ $<$ k$T_{\rm in}$). We present a possible explanation for the soft excess in the discussion.

XMMU\thinspace J004243.6+412519 is a transient X-ray source in M31, identified for the first time in a 2012 January 15 {\em XMM-Newton} observation, hereafter observation X1 \citep{henze2012a}; its  luminosity was found to be $\sim$2$\times 10^{38}$ erg s$^{-1}$. \citet{henze2012b} observed the source again twice more with {\em XMM-Newton} (observations X2--X3); the highest observed luminosity (in X3) was 2.5$\times 10^{39}$ erg s$^{-1}$, assuming a two component emission model consisting of a power law with photon index $\Gamma$ = 2.7$\pm$0.1, and a disk blackbody with  inner disk temperature k$T_{\rm in}$ = 0.79$\pm$0.05 keV, suffering line of sight absorption equivalent to 6.2$\pm$0.5$\times 10^{21}$ H atom cm$^{-2}$. This luminosity is a factor $\sim$2 higher than the Eddington limit for a standard 10 $M_{\odot}$ black hole, hence \citet{henze2012b} labeled it an ultra-luminous X-ray source. Since this is the second ULX in M31, we will refer to this source as M31 ULX2.

 \citet{henze2012c} followed up the {\em XMM-Newton} observations of M31 ULX2 with five {\em Swift} observations (S1--S5) over 2012 February 19 -- March 4 that were  consistent with constant intensity; they combined the spectra, and found the summed spectrum to be well described by either a pure 0.88$\pm$0.04 keV disk blackbody or a 0.9$\pm$0.3 keV disk blackbody + power law with $\Gamma$ fixed to 2.7. A further 2012 May 24 {\em Swift} observation (S6) resulted in a spectrum that was well described by a 0.66$^{+0.07}_{-0.06}$ keV disk blackbody, absorbed by 4.0$^{+1.4}_{-1.1}\times 10^{21}$ atom cm$^{-2}$ \citep{henze2012d}.

\citet{middleton12} discovered strong variable radio emission associated with the X-ray source during outburst,  and found it likely that the radio source was a jet powered by near-Eddington accretion onto a stellar mass black hole.

 They also carefully modeled the X-ray spectra from X1--X3, and found that an emission model consisting of a disk blackbody ({\sc diskbb}) + Comptonization model ({\sc comptt}) was  preferred  for X2 and X3 over the disk blackbody + power law model used by Henze et al. (2012b); however, they do not discuss this in detail. For their {\sc diskbb + comptt} models, k$T_{\rm in}$ $\sim$0.5 keV, $T_0$ $\sim$1.2 keV, and the optical depth of the corona was $\sim$11; they did not give the electron temperature. The resulting 0.3--10 keV luminosities were considerably lower than those obtained by Henze et al. (2012b): 9.8$\pm$0.2$\times 10^{38}$ and 1.26$\pm$0.02$\times 10^{39}$ erg s$^{-1}$ for X2 and X3 respectively. \citet{middleton12} present spectra for M31 ULX2 at various stages of its spectral evolution.

The \citet{middleton12} emission model represents the $T_0$ $>$ k$T_{\rm in}$ family of models, while the \citet{henze2012b} model represents the $T_0$ $<$ k$T_{\rm in}$ model family.  We will therefore consider the  luminosity solutions for X2 and X3 from Henze et al. (2012b), as well as those of  Middleton et al. (2012).

Two types of X-ray binaries (XBs) exhibit transient behavior. Low mass X-ray binaries may be transient X-ray sources due to instabilities in their accretion disks; the disk has two stable phases (hot and cold), and an unstable intermediate phase--- matter accumulates in the disk in the cold phase, and is rapidly dumped onto the compact object in the hot phase \citep[see e.g.][]{lasota2001}. However, the X-rays produced by accretion from the hot disk prevent the disk from cooling; the X-ray luminosity decays exponentially if the whole disk is ionized, and linearly if only part of the disk is ionized\citep {king98}. Meanwhile high mass X-ray binaries with large eccentricities may be transient if mass transfer is only possible near periastron, or if Be class donor stars experience ejection events \citep[see e.g.][]{stella86}.

We have been monitoring the central region of M31 for the last $\sim$13 years with {\em Chandra}, averaging $\sim$1 observation per month,  looking for X-ray transients. Promising examples are followed up with two {\em HST} ACS observations, the first is taken a few weeks after outburst, and the second observation is taken $\sim$6 months later; this allows us to identify the counterpart via difference imaging \citep[see e.g.][ and  references within]{barnard2012b}.

\citet{vp94} found an empirical relation between the  ratio of  X-ray and optical luminosities of Galactic X-ray binaries  and their orbital periods, suggestive that the optical emission is dominated by reprocessed X-rays in the disk; this relation appears to be independent of inclination. Their chosen X-ray band was 2--10 keV. For an irradiated accretion disk with radius $a$, X-ray luminosity $L_{\rm X}$, optical luminosity $L_{\rm opt}$, and temperature $T$, T$^4$ $\propto$ $L_{\rm X}$/$a^2$, while the surface brightness of the disk, $S$, $\propto$ T$^2$ for typical X-ray binaries \citep{vp94}. Since $L_{\rm opt}$ $\propto$ $S.a^2$,  $L_{\rm opt}$ $\propto$ $L_{\rm X}^{1/2} a$; also $a$ $\propto$ $P^{2/3}_{\rm orb}$, where $P_{\rm orb}$ is the orbital period. 

\citet{vp94} defined $\Sigma$ = $\left(L_{\rm X}/L_{\rm EDD}\right)^{1/2}\left(P_{\rm orb}/1 {\rm hr}\right)^{2/3}$, choosing   $L_{\rm EDD}$ = 2.5$\times 10^{38}$ erg s$^{-1}$ as a normalizing constant, and found 
\begin{equation}
M_{\rm V} = 1.57(\pm0.24) - 2.27(\pm 0.32) \log \Sigma.
\end{equation}
However, \citet{vp94} sampled a mixture of neutron star and black hole binaries, in various spectral states. A cleaner sample was obtained by A. Moss et al. (2013, in prep), who used only black hole transients at the peaks of their outburst, and found
\begin{equation}
M_{\rm V} = 0.84(\pm0.30) - 2.36(\pm0.30) \log \Sigma.
\end{equation}
We note that these two relations only differ significantly in normalization, caused by black hole X-ray binaries having larger disks than neutron star binaries with the same period. We have period estimates for 12 M31 transients observed by {\em Chandra} and {\em HST} \citep{barnard2012b}.

We used our first {\em Chandra} observation  of M31 ULX2 (2012 February 19, C1) to refine its X-ray position to 00:42:43.683 +41:25:18.53 in J2000 co-ordinates, with 1$\sigma$ uncertainties of 0.20$''$ in RA and 0.14$''$ in Dec \citep{barnard2012a}. We then triggered two {\em HST}/ACS observations, using the F435W filter that approximates the B band.

In this work we present analysis of our {\em Chandra} and {\em HST}  observations of M31 ULX2 (C1--C9 and H1--H2 respectively), as well as a serendipitous observation in a 120 ks {\em XMM-Newton} observation (PI R. Barnard); we refer to this observation   as X4. We also re-analyzed the {\em Swift} observations S1--S6, in order to obtain uncertainties on their fluxes; X1--X3 were not available to the public. We also give a range of orbital periods estimated for the system.

\section{Observations and data reduction}

We  observed M31 ULX2  nine times with {\em Chandra} ACIS over 2012 February -- September (C1--C9); unfortunately, M31 was unobservable for much of March and April, resulting in a $\sim$100 day delay between C1 and C2. Our two {\em HST} observations were made on 2012 March 7 (H1) and September 7 (H2).  We serendipitously observed M31 ULX2 with {\em XMM-Newton} for 120 ks on 2012 June 26 (X4, PI R. Barnard).

\subsection{Optical analysis}

All optical analysis was performed with {\sc pc-iraf} Revision 2.14.1, except where noted. Each {\em HST} observation included four flat-fielded (FLT) images, and one drizzled  (DRZ) image. The flat fielded images are corrected for instrumental effects, but not background subtracted; the total number of counts in each pixel is given. The native ACS resolution is comparable to the FWHM of the PSF \citep{fruchter09}. The drizzled image combines the flat-fielded images, removes any cosmic rays, and subtracts the sky background; it is normalized to  give the number of counts per second per pixel.   We used the DRZ images from H1 and H2 to create a difference image; however, we mainly used the FLT images for our aperture photometry because the software used  ({\sc daophot}) prefers images that include the sky background and measure  brightness  in counts. 

\subsubsection{Creating a difference image}

 We reprojected the  H2 DRZ  image  into the coordinates of the H1  DRZ image, to produce an accurate difference image. To do this, we first registered the H2 image to the H1 image with {\sc ccmap}, using unsaturated stars that were close to the target; this maximized the accuracy of the registration at the position of the target. Then, we used the {\sc iraf} task {\sc wregister} to make the pixel orientation of the H2 image match that of the H1 image. We registered the H2 image to the H1 image before mapping to reduce the noise during image subtraction.   The difference image was produced by subtracting H2 from H1 using the {\sc ftools} task  {\sc farith}. 

\subsubsection{Measuring the optical counterpart}

For our first {\em HST}  observation (H1) we used the {\sc daophot} package released with {\sc iraf} to obtain the net source counts in the FLT images, for a total of $C_{\rm tot}$ counts over $T$ seconds. In particular, we used {\sc daoedit} to obtain the number of source counts from radial profile fitting.
 We converted this  to Vega $B$ magnitude via
\begin{equation}
B \simeq -2.5 \log\left[ C_{\rm tot}/T \right] + ZP,
\end{equation} 
having obtained the zero point ($ZP$ = 25.77) from the ACS Zero Point calculator\footnote{http://www.stsci.edu/hst/acs/analysis/zeropoints/zpt.py}; we see from \citet[][Equation 12 and Table 18]{sirianni05} that the conversion from F435W counts  to B magnitude is within 3$\sigma$ of our ZP for B-V = $-$0.09 \citep[assuming a typical disk spectrum][]{liu01}.

 We can convert from $B$ magnitude to $M_{\rm V}$ via
\begin{equation}\label{conv}
 M_{\rm V} = B + 0.09   -   N_{\rm H}\times\left(1+1/3\right)/1.8\times10^{21}  -    24.47, 
\end{equation}
where $N_{\rm H}$ is the line of sight absorption; this accounts for the difference in B and V magnitudes of a typical accretion disc,  a relationship between B band extinction and measured line-of-sight absorption towards the object, and the distance to M31 \citep[see][and references within]{barnard2012b}.

For the second observation (H2) we examined each FLT image and found the total number of background counts in a circular region with 3 pixel radius at the position of the counterpart; this corresponds to $\sim$3 times the FWHM width of the PSF. We obtained the 4$\sigma$ $B$ magnitude upper limit from $4\left( C_{\rm tot}\right)^{0.5}/T$.

\begin{figure*}[!ht]
\epsscale{1}
\plotone{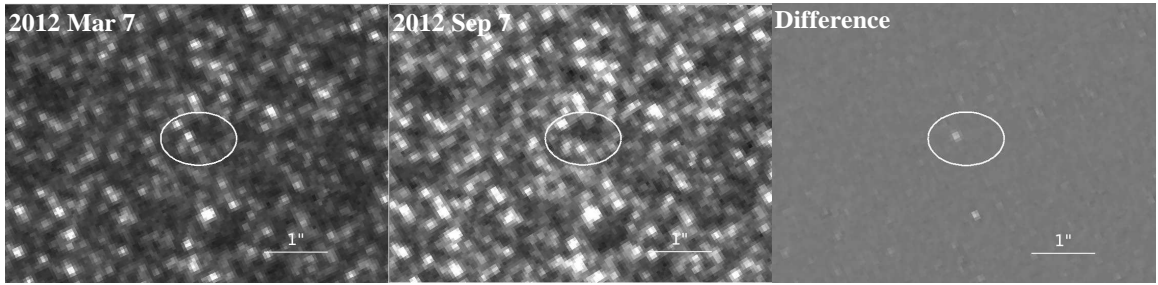}
\caption{Details of two {\em HST}/ACS observations of M31 ULX2 using the F435W filter, along with the difference image. North is up, east is left, a 1$''$ scale is indicated. The 3$\sigma$ uncertainty in the X-ray position is indicated by an ellipse. A  white object in the difference image is brighter in the first observation than in the second one; the counterpart to M31 ULX2 is easily identified.}\label{optim}
\end{figure*}

\subsection{X-ray Analysis}
We extracted spectra from our X-ray observations of M31 ULX2 using  the appropriate mission-specific software suites: CIAO v.4.4 for C1--C9,  HEASOFT v6.13 for S1--S6, and {\em XMM-Newton} SAS v.12.0.1 for X4. We used the CALDB version distributed with CIAO. Spectral analysis was performed with XSPEC v12.7.1b.

\subsubsection{{\em Chandra} analysis}
For each of our {\em Chandra} ACIS observations (C1--C9), we extracted source and background 0.3--7.0 keV spectra from circular regions with 20$"$ radii; the high off-axis angle meant that the PSF was considerably larger than for an  on-axis point source. We then created a response matrix using {\sc mkacisrmf}, and obtained an ancillary response file from {\sc mkarf}. The source spectra were grouped to give a minimum of 20 counts per bin using {\sc grppha}.

\subsubsection{{\em Swift} analysis}
Following the recommended procedure, we extracted source spectra from a circular region with 20 pixel (47$"$) radius, using {\sc xselect} v2.4b. The background spectra were accumulated from a nearby circular region with 40 pixel (94$"$) radius. We then produced ancillary response files using {\sc xrtmkarf}, using the included exposure map to account for variations in effective area over the observation. {\em Swift} response files are not generated on the fly; instead, the correct response is found by using the {\sc quzcif} tool, and copied into the working directory. Spectral channels 0--29 were labeled bad, and the spectra were grouped to give at least 20 counts per bin using {\sc grppha}.

\subsubsection{{\em XMM-Newton} analysis}
Due to the high  intensity of M31 ULX2, we only used data from the pn instrument. {\em XMM-Newton} observations often experience intervals of greatly increased background levels. We searched for such flaring intervals by creating a lightcurve with {\sc xmmselect}, using the expression ``(PATTERN==0)\&\&(PI in [10000:12000])\&\&(FLAG==0)'' and 100 s bins; we filtered out all intervals where the pn rate was $>$0.4 count s$^{-1}$ using {\sc tabgtigen}. Flaring was substantial, with 75 ks of good time out of a 120 ks total.

We extracted the source spectrum from a region that was optimized by the analysis software; this was a circle with radius $\sim$24$"$. A circular background region was chosen to be near the source, on the same chip, and at a similar off-axis angle; its radius was $\sim$40$"$. These spectra were filtered by the good time interval, and by the expression ``(PATTERN$<$=4)\&\&(FLAG==0)''. We obtained a corresponding response file with {\sc rmfgen} and an ancillary response file with {\sc mkarf}. The source spectrum was grouped to a minimum of 20 counts per bin, and energies outside the range 0.3--10 keV were excluded.

\subsection{Locating the X-ray source}
We were unable to register the {\em Chandra} position of M31 ULX2 directly with the {\em HST} observations due to a lack of known X-ray sources in the {\em HST} field. Instead
 we used 27 X-ray bright globular clusters (GCs) to register a combined $\sim$350 ks  ACIS image (supplied by Z. Li) to the B band Field 5 image of M31 provided by the Local Galaxy group Survey (LGS) \citep{massey06}. We used {\sc pc-iraf} v2.14.1 to perform the registration, following the same procedure as described in \citet{barnard2012b}. The X-ray and LGS positions of the 27 X-ray bright GCs  were determined using  {\sc imcentroid}; the equivalent FK5 coordinates were calculated for the X-ray and optical position of each GC using {\sc xy2sky} v2.0, distributed with {\sc ftools}. The X-ray positions of each GC were altered to match the LGS positions, allowing the registration of the merged {\em Chandra} image to Field 5 using the {\sc iraf} task {\sc ccmap}. This registration yielded  1$\sigma$ position uncertainties of  0.11$"$ in R.A., and 0.09$"$ in Dec \citep{barnard2012b}.

We then registered our first {\em HST}  observation (H1)  to the Field 5 observation in the same way using bright, unsaturated stars. Similarly, we registered our brightest {\em Chandra} observation of M31 ULX2 (C1) to the merged {\em Chandra} image using a selection of bright X-ray sources. The final uncertainties in the X-ray position of M31 ULX2 combine the position uncertainties in the X-ray image, the uncertainties in registering Observation C1 to the merged {\em Chandra} image, and the uncertainties in registering the merged {\em Chandra} image to the M31 Field 5 LGS image. The uncertainties in registering the {\em HST} observation were negligible.

\section{Results}
\label{res}

\subsection{Optical analysis}
\subsubsection{Difference imaging}
The difference image produced from the H1 and H2 images revealed an optical counterpart at 00:42:43.699 +41:25:18.54 (J2000), with 1$\sigma$ uncertainties of 0.21$''$ RA and 0.18$''$ Dec with respect to the LGS M31 Field 5 image. We present details of each {\em HST} observation in Fig.~\ref{optim}, along with the difference image.

\subsubsection{Photometry}

We first obtained source counts in H1 for the ULX2 counterpart from FLT images  2 and 3 using PSF fitting via {\sc daoedit}:  1 and 4 were contaminated. For Frame 2, we obtained 1010 net source counts over 1255 s; the sky mean was 141 count pixel$^{-1}$, with $\sigma$ = 33 count pixel$^{-1}$. For Frame 3, we obtained 1198 net source counts over 1410 s (sky mean = 166 count pixel$^{-1}$, $\sigma$ = 39 count pixel$^{-1}$). Individually, these frames gave a B magnitude of 26.03$^{+0.05}_{-0.03}$ and 25.95$^{+0.03}_{-0.02}$ respectively. Combining these frames yielded B = 25.97$^{+0.03}_{-0.02}$. For consistency, we also checked the B magnitude derived from the DRZ image; this yielded a source intensity of 0.867 count s$^{-1}$ over 5450 s (sky mean = $-$0.026 count s$^{-1}$ pixel$^{-1}$, $\sigma$ = 0.011 count s$^{-1}$ pixel$^{-1}$); the DRZ image gave B = 25.924$_{0.015}^{+0.017}$, consistent with the magnitude from our FLT images.
 We will use our B magnitude from the combined FLT frames (B = 25.97$^{+0.03}_{-0.02}$) for the rest of this work.

We saw no sign of the counterpart to M31 ULX2 in observation H2. Instead we estimated the 4$\sigma$ upper limit to the B magnitude from the background counts; we note that 3 pixel radius circles used may be contaminated by neighboring stars, but this is not too important since we are only interested in upper limits. FLT images 1--3 yielded a total of 6829 counts over 3861 seconds; image 4 was contaminated by a cosmic ray. Hence, B $>$ 28.4 at the 4$\sigma$ level.

The optical luminosity of the M31 ULX2 counterpart varied by at least $\sim$1 order of magnitude between observations. Using the value for $N_{\rm H}$ obtained from X4, $M_{\rm V}$ = $-$0.91 for H1, and  $>$+1.5 for H2; the counterpart was more luminous (in terms of absolute magnitude) than any other transient counterpart found in M31 to date, apart from ULX1 \citep{barnard2012b}; however, it was also fainter (in apparent magnitude) than any other counterpart due to the high absorption (see below). The known counterparts of HMXBs in the Small Magellanic Cloud have absolute values $-6$ $\la$ $M_{\rm V}$ $\la$ $-1$; for the  known BH HMXBs (Cygnus X-1, LMC X-1, and LMC X-3), $-6.5$ $\la$ $M_{\rm V}$ $\la$ $-1.5$ \citep[and references within]{barnard11}. All of these high mass systems are optically brighter than the M31 ULX2 counterpart even in outburst; hence a low-mass donor is likely.

\subsection{X-ray analysis}

We summarize the X-ray properties of M31 ULX2 during all the observations that we analyzed in Table~\ref{restab}. For each observation we give the time since the first observation, the absorption, as well as the disk blackbody temperature and/or power law photon index depending on the spectral state; the hard state was represented by a power law ({\sc xspec} model {\sc wabs*cflux*pow}) with photon index 1.7 (as is typical),  the thermally dominated state was represented by {\sc wabs*cflux*diskbb}, while the steep power law state was represented by {\sc wabs*(cflux*diskbb + cflux*pow)}. Finally we give the $\chi^2$/dof, luminosity and spectral state; luminosities assume a distance of 780 kpc \citep{sg98}. Values quoted without uncertainties are frozen. All uncertainties  in results presented in this work are quoted at the 1$\sigma$ level.

\begin{figure}[!ht]
\epsscale{1}
\plotone{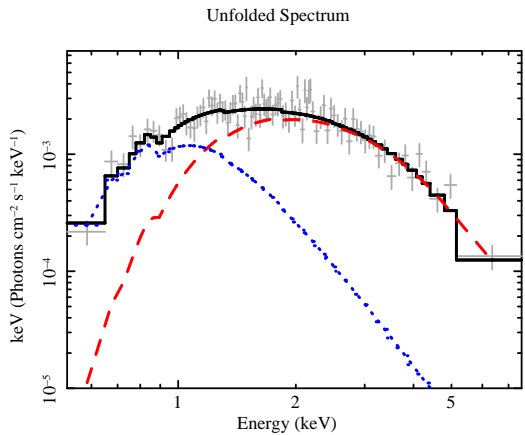}
\caption{The 0.5--10 keV S4 XRT spectrum fitted with the best Scenario B model. The disk blackbody and power law components are represented by dashes and dots respectively. A color version is provided in the electronic edition.}\label{swiftspec}
\end{figure}

The spectra of X2 and X3 have been interpreted in two ways. Henze et al. (2012b) used a disk blackbody + power law emission model, and obtained a 0.3--10 keV luminosity of $\sim$2.5$\times 10^{39}$ erg s$^{-1}$ for X3. Alternatively, Middleton et al. (2012) used a disk blackbody + compTT emission model with a 0.3--10 keV luminosity for X3 of 1.26$\pm$0.02$\times 10^{39}$ erg s$^{-1}$. We note that the fit obtained by Middleton et al. (2012) suggests a compact,  optically thick, opaque  corona, while the Henze et al. model represents an extended, optically thin corona \citep[see e.g.][]{barnard11}; we shall label these Scenario A and Scenario B respectively. 

\subsubsection{Re-analyzing the {\em Swift} data}

Following Henze et al. (2012c), we fitted S1--S5 simultaneously, with all of the parameters free to vary but tied between observations; we obtained similar results. However, we note that the number of degrees of freedom in our fits were substantially smaller than for the fits quoted by Henze et al. (2012c); this is probably due to different binning strategies. Our best simultaneous fit for a disk blackbody emission model yielded $N_{\rm H}$ = 2.87$\pm0.15\times 10^{21}$ atom cm$^{-2}$, and  k$T_{\rm in}$ = 0.877$\pm$0.016 keV; $\chi^2$/dof = 377/332. When fitting a disk blackbody + power law model, the photon index was frozen at 2.7, following \citet{henze2012b}; $N_{\rm H}$ = 3.9$^{+0.4}_{-0.5} \times 10^{21}$ atom cm$^{-2}$, k$T_{\rm in}$ = 0.85$\pm$0.02 keV, and $\chi^2$/dof = 374/331. 

The null hypothesis probabilities for the model fits to S1--S5 are 9\% for Scenario A, and 10\% for Scenario B. We caution that there is no statistical requirement for the addition of the power law component to these relatively low quality spectra; our motivation for including Scenario B  is the discrepancy between the Scenario A lightcurve and  our optical results, c.f. the  known behavior of Galactic transients (see below). We also note that our unabsorbed luminosities for Scenario B assume a particular power law index (following Henze et al., 2012bc); Scenario B luminosities may therefore be subject to systematic uncertainties; however, our X4 spectrum is consistent with $\Gamma$ = 2.7 at the 1$\sigma$ level.

The 0.3--10 keV luminosity for the disk blackbody model for S1--S5 was 1.11$\pm$0.02$\times 10^{39}$ erg s$^{-1}$, while the two-component model yielded a luminosity of 1.5$\pm$0.3$\times 10^{39}$ erg s$^{-1}$. These two luminosities are consistent, due to the large uncertainties in the luminosity from the two-component model. In physical terms, the extended corona would appear brighter because cool photons in the outer disk are up-scattered into the observed energy range. 

In Fig.~\ref{swiftspec} we present the best fit  unfolded Scenario B spectrum  multiplied by energy for S4 (the highest quality Swift spectrum); the disk blackbody and power law components are represented by dashes and dots respectively. We see that the power law component dominates below $\sim$1 keV; this is physically possible for an extended corona but not for a compact corona. The best fit $\chi^2$/dof = 98/87. Setting the power law contribution to zero increases the $\chi^2$/dof to 148/88, demonstrating that the power law contribution is significant in this model; however, a single disk blackbody emission model fits this spectrum just as well, with $\chi^2$/dof = 99/88.

Our analysis of S6 supports that analysis reported by \citet{henze2012d}. Fitting an absorbed disk blackbody model yielded $N_{\rm H}$ = 3.5$\pm$0.9$\times 10^{21}$ atom cm$^{-2}$, and k$T_{\rm in}$ = 0.62$\pm$0.05 keV; $\chi^2$/dof = 19/18 with a 0.3--10 keV luminosity of 4.8$\pm$0.6$\times 10^{38}$ erg s$^{-1}$.

\begin{table*}
\begin{center}
\caption{ Summary of  our spectral fitting results for  M31 ULX2. For each observation we give the time from first observation, absorption normalized to 10$^{21}$ atom cm$^{-2}$, disk blackbody temperature, and power law photon index; we then give the corresponding $\chi^2$/dof,  0.3--10 keV luminosity normalized to 10$^{37}$ erg s$^{-1}$, and spectral state. For C1 and S1--S5 we give fits pertaining to Scenario A (TD) and B (SPL); S1--S5 were fitted simultaneously with all parameters tied. The quality of the spectrum is indicated by the degrees of freedom (dof), since each spectral bin contains at least 20 counts. For C8--C9, we assume a hard state emission model, with $N_{\rm H}$ = 3.37$\times 10^{21}$ atom cm$^{-2}$, and $\Gamma$ = 1.7. All uncertainties are quoted at the 1$\sigma$ level.} \label{restab}
\renewcommand{\tabcolsep}{2pt}
\renewcommand{\arraystretch}{1.1}
\begin{tabular}{cccccccccccc}
\tableline\tableline
Obs &$T-T_0$& $N_{\rm H}^{21}$ & k$T_{\rm in}$ / keV & $\Gamma$ & $\chi^2$/dof& $L_{37}$ & State  \\
\tableline 
S1 &  34.4 & 2.87$\pm$0.15 & 0.877$\pm$0.016 &  --- & 377/332 & 111$\pm$2 & TD (A)\\
C1 &  34.6 & 2.4$\pm$0.5 & 0.97$\pm$0.05 &  --- & 131/127 & 106$\pm$3 &  TD (A)\\
S2 &  39.2 & 2.87$\pm$0.15 & 0.877$\pm$0.016 &  --- & 377/332 &111$\pm$2 & TD (A)\\
S3 &  46.6 & 2.87$\pm$0.15 & 0.877$\pm$0.016 &  --- & 377/332 &111$\pm$2 & TD (A)\\
S4 &  47.4 & 2.87$\pm$0.15 & 0.877$\pm$0.016 &  --- & 377/332 & 111$\pm$2 & TD (A)\\
S5 &  48.4 & 2.87$\pm$0.15 & 0.877$\pm$0.016 &  --- & 377/332 & 111$\pm$2 & TD (A)\\
S1 &  34.4 & 3.9$^{+0.4}_{-0.5}$ & 0.85$\pm$0.02 &  2.7 & 374/331 & 150$\pm$30 & SPL (B)\\
C1 &  34.6 & 4.5$^{+0.7}_{-0.9}$ & 0.94$\pm$0.04 &  2.7 & 127/126 &175$\pm$50 & SPL (B)\\
S2 &  39.2 & 3.9$^{+0.4}_{-0.5}$ & 0.85$\pm$0.02 &  2.7 & 374/331 & 150$\pm$30 & SPL (B)\\
S3 &  46.6 & 3.9$^{+0.4}_{-0.5}$ & 0.85$\pm$0.02 &  2.7 & 374/331 &150$\pm$30 & SPL (B)\\
S4 &  47.4 & 3.9$^{+0.4}_{-0.5}$ & 0.85$\pm$0.02 &  2.7 & 374/331 &150$\pm$30 & SPL (B)\\
S5 &  48.4 & 3.9$^{+0.4}_{-0.5}$ & 0.85$\pm$0.02 &  2.7 & 374/331 &150$\pm$30 & SPL (B)\\
H1 &  50.9 & --- & --- &  --- & --- & ---\\
S6 & 130.0 & 2.9$\pm$0.7 & 0.64$\pm$0.05 &  --- & 20/19 & 45$\pm$4 & TD\\
C2 & 131.9 & 5.0$\pm$1.7 & 0.62$\pm$0.07 &  --- & 14/15 & 47$\pm$8 & TD\\
C3 & 137.5 & 3.18$\pm$0.10 & 0.656$\pm$0.007 &  --- & 257/222 & 53.0 $\pm$0.7 & TD\\
C4 & 142.7 & 3.24$\pm$0.11 & 0.645$\pm$0.007 &  --- & 196/210 &48.7$\pm$0.7 & TD\\
C5 & 148.7 & 3.23$\pm$0.11 & 0.641$\pm$0.007 &  --- & 200/213 & 45.0$\pm$0.6 & TD\\
C6 & 157.8 & 3.2$\pm$1.2 & 0.61$\pm$0.05 &  --- & 28/23 & 33$\pm$5 & TD\\
X4 & 163.3 & 3.37$\pm$0.07 & 0.577$\pm$0.005 &  2.3$^{+0.4}_{-0.8}$& 684/634 & 40.7$\pm$1.0 & TD\\
C7 & 186.3 & 3.5$\pm$1.4 & 0.58$\pm$0.08 &  --- & 18/19 &  20$\pm$3 & TD\\
C8 & 216.0 & 3.37 & --- &  1.7 & N/A &2.6$\pm$0.4 & HS\\
H2 & 234.9 & --- & --- &  --- & --- & ---\\
C9 & 240.6 & 3.37 & --- &  1.7 & N/A &0.7$\pm$0.3 & HS\\

\tableline
\end{tabular}

\end{center}
\end{table*}


\subsubsection{The X-ray evolution of M31 ULX2}

 Both scenarios agree on the following. The X-ray emission of M31 ULX2 evolved considerably during the outburst: it appears to have evolved from the hard state (HS)  to a two component emission  state during the rise (X1--X3), either the ``ultra-luminous'' state or something resembling the the steep power law (SPL) state; we note that the true peak may not have been observed. It then went to the  thermally dominated (TD) state in decline (X4, C2--C7, S6). A sudden drop in intensity at the end of the outburst (C8--C9) suggests a transition from the TD state (where the most of the bolometric luminosity is in the observed band) to the HS state (where much of the power can be emitted at much higher energies).

The difference between Scenarios A and B is in the interpretation of the spectra from C1 and S1--S5.  These spectra  are consistent with being in a SPL-like  state (disk blackbody + power law spectra), or in the TD state (disk blackbody emission only). The TD spectra for C1 and S1--S5 are only consistent with Scenario A; the SPL spectra are consistent with both scenarios due to large uncertainties.

For Scenario A, the shape of the 0.3--10 keV lightcurve is more linear than exponential, losing $\sim$5$\times 10^{36}$ erg s$^{-1}$ per day. However,  Scenario B favors exponential decay, with a 0.3--10 keV e-folding time of $\sim$80 days. Since X-ray transient decay is expected to be exponential or linear depending on whether or not the disk is fully ionized, \citet{shahbaz98} studied the dependence of lightcurve shape on orbital period and peak X-ray luminosity. For black hole transients with peak X-ray luminosities $>$10$^{39}$, they expect exponential decay even for systems with periods $>$300 hr. For Scenario A to be correct,  the optical counterpart is expected to have $M_{\rm V}$  $\la$ $-5$ according to either Equation 1 or 2, and $B$ $\la$21 according to Equation 4; this is considerably brighter than the observed magnitude, $B$ = 25.95$^{+0.03}_{-0.02}$, hence we find Scenario A unlikely.


\begin{figure}[!b]
\epsscale{1}
\plotone{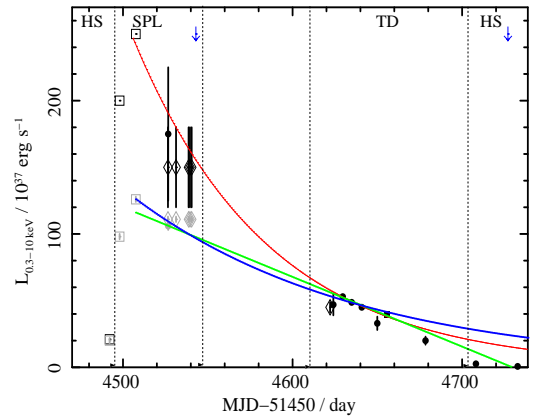}
\caption{Unabsorbed luminosity lightcurve of M31 ULX2 in the 0.3--10 keV band. Black open squares represent {\em XMM-Newton} luminosities taken from Henze et al (2012ab); no uncertainties were given. Grey open squares represent the luminosities for X1--X3 obtained by \citet{middleton12}. The re-analyzed {\em Swift} observations are represented by open diamonds.  Our {\em Chandra} ACIS observations are shown as filled circles, and our {\em XMM-Newton} observation is represented by a filled square; uncertainties are quoted at a 1$\sigma$ level. For C1 and S1--S5,  gray  and black  symbols represent  spectra for Scenarios A (TD) and B (SPL) respectively. {\em HST} observation times are indicated by downward arrows.  Vertical lines split the lightcurve by inferred spectral state: hard state (HS), thermally dominated (TD), and steep power law (SPL), following \citet{remillard06}.  The green and blue lines represent linear and exponential respectively for Scenario A, while the red line represents exponential decay for Scenario B. }\label{lc}
\end{figure}

We present a lightcurve of 0.3--10 keV luminosities for M31 ULX2 in Fig.~\ref{lc}, created using the results of Henze et al. (2012ab) and Middleton et al. (2012) in addition to our {\em Chandra}, {\em Swift}, and {\em XMM-Newton} results. We present only one datum for observations that both scenarios agree on: C2--C9, X4, and S6. Grey and black data correspond to Scenarios A and B respectively for X1--X3, C1, and S1--S5. X1--X3 are represented by open squares, X4 by a filled square, C1--C9 by circles, and S1--S6 by diamonds.  Times of  H1 and H2  are indicated by downward arrows.

\begin{figure}[b!]
\epsscale{1}
\plotone{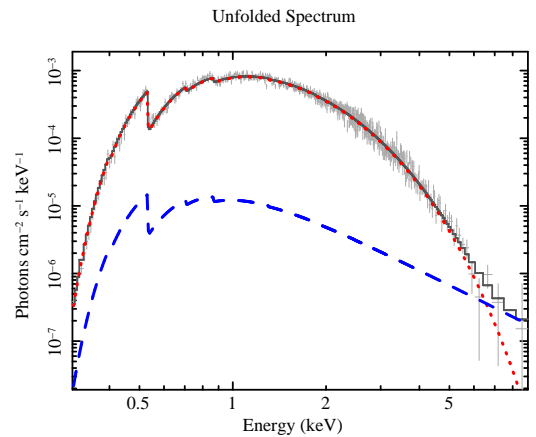}
\caption{Unfolded 0.3--10 keV {\em XMM-Newton} pn spectrum of M31 ULX2 from 2012 June 26, fitted with a disk blackbody (red dotted line) and a power law component (blue dashed line). The disk blackbody component contributes $\sim$97\% of the total 0.3--10 keV emission. A color version is provided in the electronic edition.}\label{ulx2spec}
\end{figure}

We used our X4  0.3--10 keV pn  spectrum of M31 ULX2 to get the best measurement of the absorption; we obtained $\sim$62,000 net source counts. An absorbed disk blackbody provided an acceptable fit ($\chi^2$/dof = 689/636). This was slightly improved by adding a power law component ($\chi^2$/dof =684/634):  absorption  $N_{\rm H}$ = 3.37$^{+0.07}_{-0.05}\times 10^{21}$ atom cm$^{-2}$, inner disk temperature k$T_{\rm in}$ = 0.577$\pm$0.005, photon index $\Gamma$ = 2.3$^{+0.4}_{-0.8}$, quoting  1$\sigma$ uncertainties. The total 0.3--10 keV unabsorbed luminosity was 4.07$\pm$0.10$\times 10^{38}$ erg s$^{-1}$, with 3.95$\pm$0.04$\times 10^{38}$ erg s$^{-1}$ contributed by the disk blackbody; this corresponds to the TD state \citep{remillard06}. We present this best fit to the pn spectrum in Fig.~\ref{ulx2spec}.

While most of the absorption measurements are consistent with that from X4, we note that the spectral fits obtained by \citet{henze2012b} for X2 and X3 suggest significantly higher absorption.  \citet{henze2012b} use a power law component to represent the Comptonized emission component; this could overestimate the number of soft photons in the unfolded spectrum, thereby increasing the absorption required to produce observed spectrum. However, observations S1--S5, and C1 were also modeled with the same power law component in Scenario B, and yielded $N_{\rm H}$  values consistent with X4. Furthermore, the absorption obtained by \citet{middleton12} for X2 is 8$\sigma$ higher than our absorption for X4. Hence the increase in absorption during outburst is likely to be real; this extra absorbing material may have been ejected during the outburst, and dissipated by the time of X4.

 We also note that the absorption measured in the simultaneous fitting of S1--S5 in Scenario A is 3.0$\sigma$ lower than that measured in X4, further evidence against Scenario A. 



\subsection{Estimating the orbital period}

\subsubsection{Scenario A}
In Scenario A, the 0.3--10 keV luminosity of M31 ULX2 at the time of H1 was $\sim$9.7$\times 10^{38}$ erg s$^{-1}$, assuming a decrease of $\sim$5$\times 10^{36}$ erg s$^{-1}$ day$^{-1}$. Assuming the C1 TD spectrum (Table 1), this corresponds to a 2.0--10 keV luminosity $\sim$4.5$\times 10^{38}$ erg s$^{-1}$. Since M31 ULX2 is in the TD state for Scenario A, we may use Equation 2 to estimate the orbital period. The suggested orbital period is 11$\pm$8 hr at the 1$\sigma$ level, including uncertainties in the relation; this is clearly shorter than the $>$300 hr expected from the linear decay \citep{shahbaz98}.

\subsubsection{Scenario B}
For Scenario B  we estimated the 2--10 keV luminosity of M31 ULX2 at the time of H1 to be $\sim$4.4$\times 10^{38}$ erg s$^{-1}$.  Using the relation specific to black holes, we obtained an orbital period of 8.5$\pm$0.4 hr, not including uncertainties in the relation.

 However, we note that M31 ULX2 is likely to have been in the steep power law state during H1, whereas the black hole period relation (Equation 2) is calibrated for thermally dominated spectra; the X-ray to optical ratio is likely to be different for different spectral states in the same system. Using Equation 1 results in an orbital period of 28.5$\pm$1.3 hr; given that the sample used in deriving Equation 1 was dominated by neutron stars, and that black hole binaries with the same size disk have shorter periods than neutron star binaries, we consider this period to be some sort of  upper limit.

 A period for M31 ULX2 in the range $\sim$9--30 hr is consistent with the period estimated for the first ULX in M31 (T9 in Barnard et al., 2012b, P = 18$^{+5}_{-6}$ hr).

\section{Discussion and conclusions}

XMMU\thinspace J004243.6+412519 (M31 ULX2) is the second ULX to be observed in M31; both of these systems are transient. We have combined analysis of our {\em Chandra}, {\em HST}, and {\em XMM-Newton} observations of M31 ULX2 with the reanalysis of {\em Swift} data and  reported results of Henze et al. (2012abcd) and Middleton et al. (2012) to build a picture for the system.

We have identified an optical counterpart by its change in B magnitude from 25.97 to $>$28.4 (M$_{\rm V}$ from $-$0.91 to $>$+1.5). Hence the optical emission is dominated by reprocessed X-rays from the disk. Since even the peak optical emission is fainter than known HMXB counterparts, a low mass donor is expected. We estimate an orbital period of $\sim$9--30 hr; this period is comparable with that estimated for the first M31 ULX. 

Two solutions were presented for fitting the spectra from the highest luminosity observations. Middleton et al. (2012) used a disk blackbody + Comptonization model that represented a typical ``ULX state'', with a cold, optically thick corona that obscures the inner regions of the disk \citep{gladstone09}; we refer to this as Scenario A. Henze et al. (2012b) used a disk blackbody + power law model that represents a hot, optically thin corona that extends over much of the accretion disk \citep{barnard10}; we refer to this as Scenario B. Scenario A results in a linear decay lightcurve, while Scenario B yields an exponential decay lightcurve. For X-ray luminosities $>$10$^{39}$ erg s$^{-1}$, linear decay is only expected for systems with periods $\gg$300 hr \citep{shahbaz98}. Since our orbital period estimate is $\sim$9--30 hr, we favor Scenario B for this system. However, we caution that there is no statistical evidence in the X-ray spectra that Scenario B is superior. Furthermore,  the Scenario B luminosities assume $\Gamma$=2.7, so the luminosities may be subject to systematic uncertainties.

Since this work provides further support for spectral models where the Comptonized component dominates the low energy emission \citep[on top of e.g.][]{barnard10,barnard11}, we consider how this could be possible. We are assuming unsaturated, inverse Compton scattering of cool photons on hot electrons in the corona; the mean number of scatterings experienced by the photons is much less than the number of scatterings required to bring the photons up to the electron energy \citep[see e.g.][and references within]{sunyaev80}.  If the soft excess is due to Comptonization of disk photons, the lower energy photons must be scattered more often than the high energy photons. This could be achieved if the corona density decreased in the inner disk region due to increased radiation pressure; the high energy photons in the inner disk would experience fewer scatterings than the low energy photons in the outer disk.

The spectral evolution of M31 ULX2 was unusual in that it went from something like the SPL state  to the TD state during the decay. However, we note that the spectrum reported for X3 (the observed peak) by \citet{henze2012b} is similar to the spectra exhibited by NGC253 ULX1, which in turn resembles the black hole + Wolf-Rayet binary IC10 X-1 \citep{barnard10}. The properties of those systems suggested extended coronae, and we proposed that many ULXs could be explained by stellar mass black hole systems with truly super-Eddington accretion overall, while the local accretion rate remained sub-Eddington. It is intriguing that in all of the observations where M31 ULX2 is super-Eddington, its spectra were reminiscent of the SPL state. However M31 ULX2 does appear to be consistent with a low mass X-ray binary containing a $\sim$10 M$_{\odot}$ black hole.

It is interesting to compare the outbursts of the two transient ULXs in M31.
The outburst of the first transient ULX  appears to have been more traditional, going from the HS to the TD state during the rise, then to the SPL and back to the HS state during decay; indeed it even exhibited a TD state at a 0.3--10 keV luminosity of 1.75$\pm$0.07$\times 10^{39}$ erg s$^{-1}$ \citep{nooraee12}. M31 ULX1 is also a low mass X-ray binary, but the mass of the black hole appears to be somewhat higher than usual.



\acknowledgments
We thank the anonymous referee for their thoughtful comments that substantially improved this paper. We thank Z. Li for merging the {\em Chandra} data. This research
has made use of data obtained from the {\em Chandra} data archive,
and software provided by the {\em Chandra} X-Ray Center (CXC).
We also include analysis of data from {\em XMM-Newton}, an ESA
science mission with instruments and contributions directly
funded by ESA member states and the US (NASA). R.B. is
funded by {\em Chandra} grants GO2-13106X and GO1-12109X,
along with {\em HST} grants GO-11833 and GO-12014. M.R.G. and S.S.M are partially supported by NASA grant NAS-03060.





{\it Facilities:} \facility{CXO (ACIS)} \facility{{\em HST} (ACS)} \facility{XMM (pn)} \facility{{\em Swift} (XRT)}.








\end{document}